\documentclass[aps,prl,twocolumn, superscriptaddress]{revtex4-1}

\usepackage[usenames,dvipsnames,svgnames,table]{xcolor}
\usepackage{graphicx}
\usepackage[dvipsnames]{xcolor}
\usepackage{amsmath}
\usepackage{blindtext}
\usepackage{printlen}
\usepackage{hyperref}
\usepackage{braket}
\usepackage{todonotes}
\usepackage[pass]{geometry}
\uselengthunit{in}
\usepackage{soul}
\usepackage{CJK}
\usepackage{float}

\newcommand{\subref}[2]{\hyperref[#1]{\ref*{#1}(#2)}}

\definecolor{refblue}{HTML}{2E2E91}
\hypersetup{colorlinks=true, citecolor=refblue, urlcolor=refblue, linkcolor=refblue}

\begin{document}
\begin{CJK*}{UTF8}{gbsn}

\title{\textit{In situ} imaging of the thermal de Broglie wavelength in an ultracold Bose gas}

\author{Jinggang Xiang}
\author{Enid Cruz-Col\'{o}n}
\affiliation{Department of Physics, Massachusetts Institute of Technology, Cambridge, MA 02139, USA}
\affiliation{Research Laboratory of Electronics, Massachusetts Institute of Technology, Cambridge, MA 02139, USA}
\affiliation{MIT-Harvard Center for Ultracold Atoms, Cambridge, MA 02139, USA}

\author{Candice C.~Chua}
\affiliation{Research Laboratory of Electronics, Massachusetts Institute of Technology, Cambridge, MA 02139, USA}
\affiliation{MIT-Harvard Center for Ultracold Atoms, Cambridge, MA 02139, USA}

\affiliation{\makebox{Department of Chemistry and Chemical Biology, Harvard University, Cambridge, MA 02138, USA}}

\author{William\CJKkern~R.~Milner}

\author{Julius de Hond}
\altaffiliation{Present address: Pasqal SAS, Fred. Roeskestraat 100, Amsterdam, Netherlands}
\author{Jacob F.~Fricke}
\altaffiliation{Present address: Institute for Quantum Electronics and Quantum Center, ETH Z\"{u}rich, 8093 Z\"{u}rich, Switzerland}

\author{Wolfgang Ketterle}

\affiliation{Department of Physics, Massachusetts Institute of Technology, Cambridge, MA 02139, USA}
\affiliation{Research Laboratory of Electronics, Massachusetts Institute of Technology, Cambridge, MA 02139, USA}
\affiliation{MIT-Harvard Center for Ultracold Atoms, Cambridge, MA 02139, USA}

\begin{abstract}
We report the first \textit{in situ} observation of density fluctuations on the scale of the thermal de Broglie wavelength in an ultracold gas of bosons. Bunching of  $^{87}$Rb atoms in a quasi two-dimensional system is observed by single-atom imaging using a quantum gas microscope. Compared to a classical ensemble, we observe a $30$ percent enhancement of the second-order correlation function. We show the spatial and thermal dependence of these correlations. The reported method of detecting \textit{in situ} correlations can be applied to interacting many-body systems and to the study of critical phenomena near phase transitions.

\end{abstract}

\maketitle
\end{CJK*}

\begin{figure}
    \centering
    \includegraphics[width=0.45\textwidth]{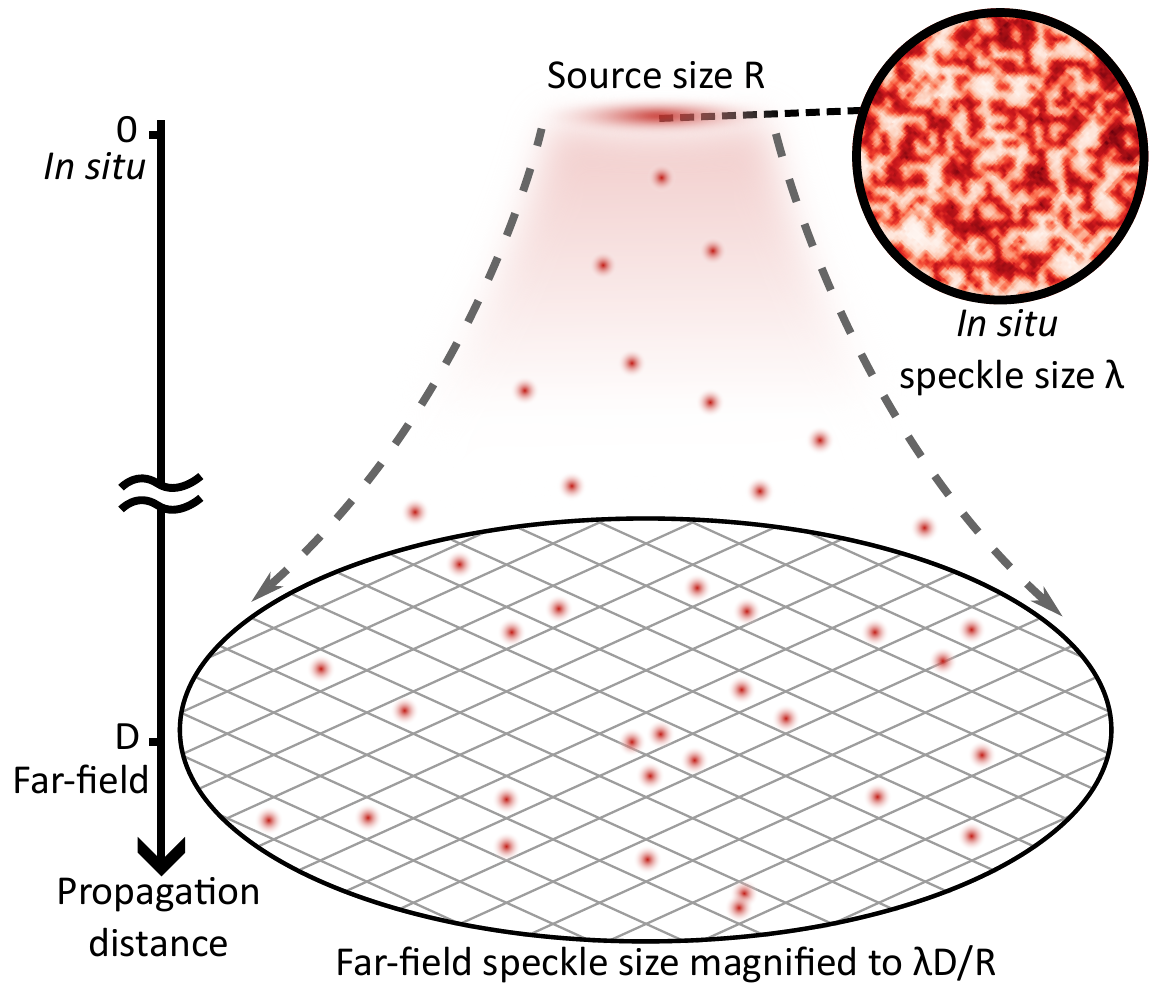}
    \caption{\textit{In situ} and far-field speckle patterns. An ensemble of identical bosons exhibits a spatial length scale $\lambda$ where density fluctuations are correlated. In photonic systems, the length scale of this speckle pattern is given by the optical coherence length, while for a gas of thermal atoms, it is the thermal de Broglie wavelength. In both the HBT experiment and previous cold atoms experiments, a free expansion over a distance $D$ is employed to enlarge the far-field speckle size by $D/R$. In this work, we use a microscope with a spatial resolution smaller than the thermal de Broglie wavelength to directly image the atomic speckle \textit{in situ}.}
    \label{fig:hbt_illustration}
\end{figure}

\paragraph*{Introduction.} 
The original Hanbury Brown and Twiss (HBT) experiment introduced the importance of correlations in photon detection and triggered the development of quantum optics~\cite{HBT1, HBT2, glauber_coherent_1, glauber_coherent_2}. HBT observed a spatial correlation of the intensity fluctuations for light emitted from the distant star Sirius. These fluctuations represent an optical speckle pattern with a characteristic length scale of the optical wavelength $\lambda$ on the star's surface. As illustrated in Fig.~\ref{fig:hbt_illustration}, when light propagates from a distant star of radius $R$ to the earth at distance $D$, the speckle size is magnified to $\lambda D/R$, which HBT determined to be around $5~\mathrm{m}$ for Sirius.

Matter waves have an equivalent speckle pattern with a characteristic grain size of the thermal de Broglie wavelength $\lambda_{\text{dB}} = h/\sqrt{2\pi m k_{B} T}$. The atomic speckle is a quantum effect, given that its scale involves Planck's constant, $h$. In contrast, a classical gas has purely Poissonian fluctuations, which has no length scale since there are no correlations in the gas. Similar to light propagation, ballistic expansion of ultracold atoms magnifies the \textit{in situ} correlation length in the far-field by $D/R$~\cite{gomes2006theory}.

Consequently, HBT experiments have been extended beyond photons to include electrons~\cite{oliver1999hanbury,henny1999fermionic,kiesel2002observation}, neutrons~\cite{iannuzzi2006direct}, cold atoms~\cite{yasuda1996observation,dall2011observation,schellekens2005hanbury,jeltes2007comparison,manning2010hanbury,manning2013third,thomas2024nbody,esteve2006observations,perrin2012hanbury,sunami2024detecting,folling2005spatial,blumkin2013observing,rom2006free,sanner2010suppression,muller2010local,guarrera2011observation}, and cold molecules \cite{rosenberg2022observation}. As a result of limited detector resolution, the atomic speckle pattern has primarily been probed in the far-field after time-of-flight expansion. \textit{In situ} studies have only probed enhanced density fluctuations on a length scale much larger than the de Broglie wavelength~\cite{esteve2006observations} or focused on temporal correlations~\cite{guarrera2011observation}. For cold fermions in a lattice, Pauli antibunching between next-neighbor sites has been observed in ~\cite{cheuk2015quantum,omran2015microscopic,hartke2020doublon}. However, the counterpart of bosonic bunching is less studied.

In this Letter, we report the first \textit{in situ} observation of the atomic correlation length in a bulk, thermal Bose gas. The correlations and speckle pattern are quantitatively described by the second-order correlation function, $g^{(2)}$, which gives the joint probability of detecting two particles. The $g^{(2)}$ function is measured by cooling a small quasi two-dimensional ensemble of approximately 100 rubidium atoms to below $10~\mathrm{nK}$, with a corresponding thermal de Broglie wavelength $\lambda_{\text{dB}} \approx 2.3~\mu\mathrm{m}$, larger than the $a_{\text{lat}}= 532~\mathrm{nm}$ pinning lattice spacing used for imaging. Detection of such sparse samples is enabled by using a quantum gas microscope with single-atom resolution. Herein, we show a clear bunching signal and observe its spatial extent to be the thermal de Broglie wavelength.

\begin{figure}
    \centering
    \includegraphics[width=0.45\textwidth]{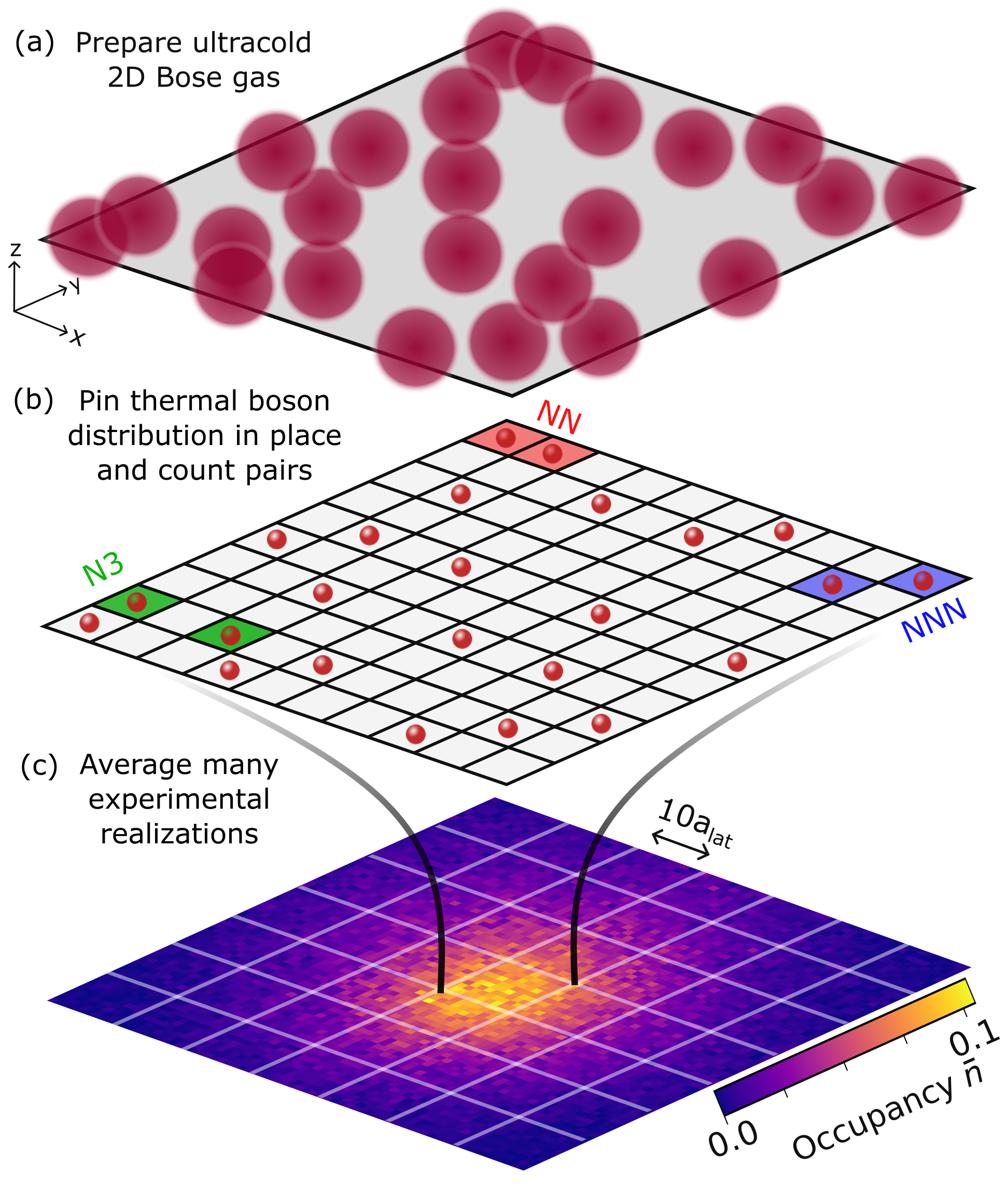}
    \caption{Experimental overview. (a) Using a $^{87}$Rb quantum gas microscope, we probe \textit{in situ} the density fluctuations in an ultracold, quasi two-dimensional thermal Bose gas. (b) We project the 2D Bose distribution onto a pinning lattice and reconstruct the site-by-site lattice occupation. To determine the bosonic enhancement, we count pairs of atoms separated by distance $r$ and determine the enhancement factor $g^{(2)}(r)$ over the classical distribution corresponding to fully distinguishable particles. On the lattice grid, we show the distance between three types of pairs, corresponding to the nearest-neighbor (NN, red), next-nearest-neighbor (NNN, blue), and third-nearest-neighbor (N3, green). (c) To experimentally determine $g^{(2)}(r)$ with high signal-to-noise, we average over 650 experimental snapshots. The region of interest is sectioned into $10 \times 10$ boxes to address the spatial inhomogeneity of the harmonic trap. The mean occupancy $\bar{n}$ on each lattice site averaged over all snapshots is plotted. We remain dilute ($\bar{n} < 0.15$) to avoid double occupation and to ensure that we do not cross the BEC phase transition. 
    }
    \label{fig:illustration}
\end{figure}

\paragraph*{Experimental setup.}  In this work, we extend quantum gas microscopy to a harmonically trapped bulk gas. This allows high quantum efficiency readout of atomic positions that far exceeds the signal-to-noise afforded by standard \textit{in situ} absorption imaging. Our experimental sequence begins by loading $^{87}$Rb atoms into a vertical lattice crossed with a horizontal trapping beam. The vertical lattice is created by focusing two 1064-nm beams onto the atoms which intersect at 5$^{\circ}$, resulting in a lattice period of 12~$\mu\mathrm{m}$. This large spacing allows the atoms to be loaded into a single layer of the vertical lattice. The combined trap is highly anisotropic, crucial for preparing a quasi two-dimensional sample. Forced evaporative cooling is applied by gradually lowering the trap depths, resulting in final trap frequencies of $(\omega_{x}, \omega_{y},  \omega_{z}) = 2 \pi \times (12, 15, 380)$~Hz. By fitting the density distribution, averaged over hundreds of experimental images, to a Bose-Einstein distribution in a harmonic trap, we determine the final temperature to be $6.5~\mathrm{nK}$.

When bosons are cooled below the phase transition temperature $T_\mathrm{c}$, they form a Bose-Einstein condensate (BEC). In analogy to single-mode lasers, a BEC has $g^{(2)}(r) = 1$, identical to that of uncorrelated classical particles. This complicates the measurement of the second-order correlation function of the Bose gas due to the spatial overlap of the BEC and thermal gas. To isolate correlations in the thermal gas, we carefully control $T_\mathrm{c}<T$ by adjusting the total atom number $N$.

During detection, we quench on a two-dimensional pinning lattice to project the bulk atomic density distribution into a square grid with 532-nm spacing. The correlation function is correspondingly discretized. Subsequently, we ramp down the vertical lattice and ramp up a light sheet for tight vertical confinement during imaging. We employ polarization gradient cooling and collect the fluorescence photons. By reconstructing fluorescence images with high fidelity, we can determine the occupancy of individual lattice sites as depicted in Fig.~\ref{fig:illustration}. Further experimental details can be found in the Supplemental Materials~\cite{Supplemental}.

\paragraph*{Analyzing correlations.}

The second-order correlation function $g^{(2)}(\textbf{r};\textbf{r}^\prime)$ can be understood as the probability of jointly detecting two particles at positions $\textbf{r}$ and $\textbf{r}^\prime$. For identical bosons at temperatures $T>T_\mathrm{c}$, where the ensemble is away from the quantum critical region, the second-order correlation function has a Gaussian form that depends on the separation distance $r = \left| \textbf{r}-\textbf{r}^\prime\right|$ between two particles~\cite{naraschewski1999spatial}:
\begin{equation}
\label{g2_eq_ideal}
    g^{(2)}(r) = 1 + \exp\left(-2 \pi r^2/\lambda_{\text{dB}}^2\right) . 
\end{equation}
When two particles occupy the same position, $g^{(2)}(0) = 2$, indicating a two-fold increase in the probability of double occupation. When pairs of atoms are sufficiently far apart that there is an absence of density-density correlation between them, $g^{(2)}(r)$ decays to 1. The spatial extent of the bosonic enhancement can be characterized by the root mean square (rms) width of the Gaussian profile, $l= \lambda_{\text{dB}}/(2 \sqrt{\pi})$, which increases as the temperature of the ensemble decreases. At our measured temperature of $6.5~\mathrm{nK}$, the rms width $l$ is $1.23~a_{\text{lat}}$.

To statistically evaluate the bosonic enhancement, we use a simple and robust analysis method: counting pairs of atoms separated by distance $r$. The measured separation between particles is discrete due to the underlying lattice structure.  The three closest types of pairs are shown in Fig.~\subref{fig:illustration}{b}, with pair separations of 1, $\sqrt{2}$, and 2 $a_\mathrm{lat}$, respectively. When the distance between atoms on two distinct lattice sites is within the thermal de Broglie wavelength, there will be an enhancement in the number of detected pairs compared to the classical expectation value.

For particles on a lattice site with indices ($i$, $j$), the nearest-neighbor (NN) pair corresponds to sites ($i \pm 1$,~$j$) and ($i$, $j\pm1$). In a homogeneous region of interest of $m \times m$ sites, assuming the probability of detecting one particle per site to be $p$, the expectation value for the number of classical (distinguishable) NN pairs is $2 m (m-1)p^2$. The $-1$ accounts for the fact that the region of interest is finite and the correction varies as the pair separation increases. For indistinguishable particles, the expectation value for the number of NN pairs is $g^{(2)}(r = a_{\text{lat}})\times 2 m (m-1)p^2$. In the absence of technical corrections, the ratio of measured pairs to classically expected pairs gives the second-order correlation function $g^{(2)}(r)$.

Ideally, we would probe the $g^{(2)}$ function at $r=0$, where the correlation signal is strongest. However, due to light-assisted collisions characteristic of quantum gas microscopes, double occupation on a single lattice site is parity projected to an empty site. Detecting pairs at $r = 0$ requires doublon detection schemes~\cite{gross2021quantum, preiss2015quantum, hartke2020doublon, koepsell2020robust}, which we do not implement in this work. The smallest separation we probe is $r = a_{\text{lat}}$ and the largest is $r \approx 5~\mu\mathrm{m}$.

\begin{figure}[t]
    \centering
    \includegraphics[width=0.49\textwidth]{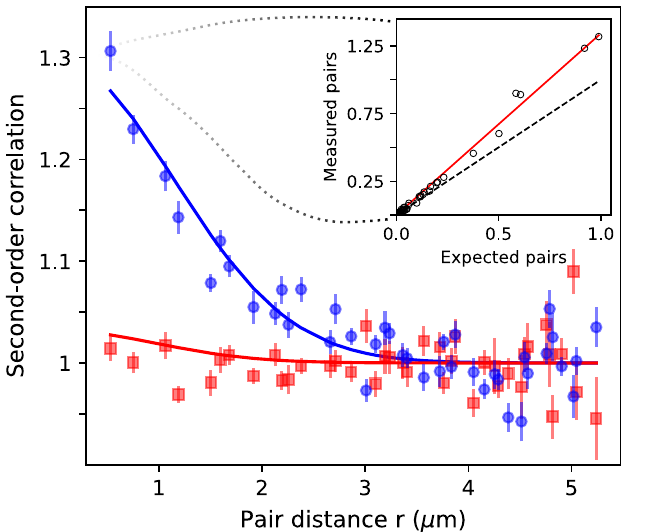}
    \caption{Spatial and thermal dependence of the second-order correlation function. $g^{(2)}(r)$ correlations determined from $\textit{in situ}$ microscope images are plotted for ensembles at temperatures $T = 6.5~\mathrm{nK}$ (blue circle) and $54~\mathrm{nK}$ (red square), respectively. The solid lines are fits using Eq.~(\ref{g2_eq}), including corrections for higher motional $n_{z}$ states and pinning lattice blurring. The thermal de Broglie wavelength is a fixed parameter. Both datasets are normalized by the mean of measurements at distance $r > 3 \; \mu$m, where no correlations are expected. Inset: For nearest-neighbor (NN) pairs, the number of measured pairs in $10 \times 10$ regions (see Fig.~\subref{fig:illustration}{c}) are plotted against the classical expectations. The red solid line is a linear fit to the data, while the black dashed line is the classical expectation. The fitted slope directly gives $g^{(2)}(r = a_\text{lat})$, and the standard deviation of the fitted slope is used as the error bar in the main plot.}
    \label{fig:correlations}
\end{figure}

Since the quasi two-dimensional cloud is prepared in a harmonic trap, the probability of detecting one particle per site has spatial dependence. We divide the total field of $70 \times 70$ sites into $49$ smaller boxes of $10 \times 10$ sites, as illustrated in Fig.~\subref{fig:illustration}{c}. This is directly analogous to applying the local-density approximation, where locally homogeneous regions are sampled. Within a $10 \times 10$ box, there is a residual spatial inhomogeneity that grows quadratically to $\approx$ 10\% for pairs separated by 5 $\mu$m. We determine the magnitude of this correction from a numerical simulation of our experimental density~\cite{Supplemental} and experimentally confirm this correction with a high-temperature measurement where correlations are minimal.  

Temporal atom number fluctuations between experimental snapshots are also a noise source that will enhance the number of observed pairs. This effect does not depend on the pair separation distance $r$ and only adds an overall scaling factor to the absolute value of $g^{(2)}(r)$. While this effect can be mitigated by post-selecting data based on the total atom number~\cite{rosenberg2022observation}, instead, we choose to include all data but normalize $g^{(2)}(r)$ by the pairs separated at a distance larger than $3~\mu\mathrm{m}$, where density fluctuations are uncorrelated.

\paragraph*{Results.}

We implement our pair analysis technique to determine the spatial second-order correlation for our thermal Bose gas, presented in Fig.~\ref{fig:correlations}. At each discrete pair separation, we evaluate the pair enhancement. For a sample prepared at $6.5~\mathrm{nK}$, we observe a $31(2)\%$ increase in the second-order correlation function for nearest-neighbor pairs separated by $a_{\text{lat}}$, as plotted in the Fig.~\ref{fig:correlations} inset. The spatial dependence matches the theoretical prediction with the thermal de Broglie wavelength $\lambda_{\text{dB}}$ being $2.3~\mu\mathrm{m}$. In contrast, when we prepare a cloud at a higher temperature of 54 nK, the observed second-order correlations are strongly diminished. 

When extrapolated to $r = 0$, the observed correlation function $g^{(2)}_{\text{obs}}(r)$ for the cold sample approaches a value of 1.3, rather than the expected value of 2. This can be attributed to the finite spatial resolution of the pinning lattice and the fact that the system is not truly two-dimensional. Our systematic understanding of this reduction can be described by the following function:
\begin{equation}
\label{g2_eq}
    g^{(2)}_{\text{obs}}(r) = 1 + \eta \frac{l^2}{\tilde{l}^2}\exp\left[-\frac{r^2}{2\tilde{l}^2}\right] . 
\end{equation}
\noindent In this expression, $\eta$ accounts for effects that display a global reduction of $g_{\text{obs}}^{(2)}(r) -1 $ (independent of $r$). The finite spatial resolution $\sigma$ of the underlying pinning lattice~\cite{gomes2006theory} broadens the correlation length $l$ to  $\tilde{l} = \sqrt{l^2+(\sqrt{2}\sigma)^2}$.

When using a pinning lattice to probe free-space atoms, the spatial resolution is not limited by the optical resolution, as long as high-fidelity image reconstruction is achieved under the quantum gas microscope. However, due to the discretized nature of the pinning lattice, the atom's position cannot be resolved beyond the lattice spacing. Consequently, the observed density distribution is the convolution of the true continuous density distribution with the lattice-imposed spatial resolution~\cite{gomes2006theory,schellekens2005hanbury}. This effect broadens the profile of $g_{\text{obs}}^{(2)}(r)$ and lowers its value at $r = 0$.

When atoms are projected onto a lattice with period $a_{\text{lat}}$, the point spread function of a simple classical model is a boxcar function with a full width at half maximum (FWHM) of $a_{\text{lat}}$, which corresponds to a rms width of $\sigma_{\text{classical}} = a_{\text{lat}}/\sqrt{12}$. For the distance between two atoms, the instrumental function is the convolution of this box with itself, producing a triangular function of FWHM $a_{\text{lat}}$ (rms width $a_{\text{lat}}/\sqrt{6}$). We model our resolution as a Gaussian with an effective rms width $\sigma$. The rms length of the measured second-order correlation function is broadened to $\tilde{l} = \sqrt{l^2+(\sqrt{2}\sigma)^2}$, where the $\sqrt2$ factor reflects the spatial resolution for detecting two-particle correlations. We fit our data using Eq.~(\ref{g2_eq}), where we fix $\eta$ from independent measurements and use only $\sigma$ as an adjustable parameter.  We obtain $\sigma = 1.25~a_{\text{lat}}$. The additional broadening compared to the classical limit is likely caused by the dynamics during the fast lattice ramp, as this can cause atoms to be excited to higher bands and have significant tunneling rates~\cite{verstraten2024situ, pyzh2019quantum}. It is an interesting question for future research to determine the ultimate quantum point spread function for a quantum gas microscope.

Even with perfect spatial resolution, the amplitude of $g_{\text{obs}}^{(2)}(r)$ can be reduced by line-of-sight integration, as we are probing a quasi two-dimensional sample. This effect does not broaden the profile of $g_{\text{obs}}^{(2)}(r)$ and is represented by the $\eta$ factor in Eq.~(\ref{g2_eq}). In a purely two-dimensional gas, where only the ground state of the vertical trapping potential is populated, this effect would be eliminated. Under our experimental conditions ($\omega_z = 2\pi\times 380~\text{Hz}$, $T = 6.5~\text{nK}$), approximately 95\% of atoms are in the ground state of the vertical trap, resulting in a contrast reduction by a factor of $ \eta = 0.91$. This reduction occurs because atoms in different  $n_z$ states are ``distinguishable" and therefore uncorrelated. 

Another fundamental reduction of the contrast is due to the effect of finite particle numbers~\cite{Supplemental, wright2012two}. As the enhanced correlations are due to exchange terms, they are reduced by $(N-1)/N$, as there is no exchange of a boson with itself.  More importantly, near the phase transition, the lowest states are already populated with multiple atoms, and this reduces the exchange terms further. For a two-dimensional box potential with $N \approx 100$, we estimate the contrast reduction to be $\approx 7\%$~\cite{Supplemental}. This effect does not significantly change our conclusions and, therefore, is not included in the fit. The effect becomes more pronounced with fewer atoms or closer proximity to the phase transition.

We also observe the temperature dependence of the second-order correlation function. By raising the vertical lattice non-adiabatically to heat the cloud, we measure a higher temperature of $54~\mathrm{nK}$, corresponding to a thermal de Broglie wavelength $\lambda_{\text{dB}}$ of $0.81~\mu\mathrm{m}$ and an rms width $l$ of $0.43~a_{\text{lat}}$. As shown in Fig.~\ref{fig:correlations}, the enhancement of the second-order correlation is substantially reduced and no longer statistically significant. The fact that the function is flat at $g_{\text{obs}}^{(2)}(r) = 1$ confirms our normalization procedure. 

\paragraph*{Outlook.}

We have directly probed the bosonic correlation length \textit{in situ} for the first time through the measurement of the second-order correlation function. We report a clear enhancement compared to the classically expected value and discuss key sources of contrast reduction. Future work can be extended in many different directions. With a homogeneous potential, one could observe modifications of the correlation function near the BEC phase transition, where the correlation length diverges and a power law decay is expected~\cite{lu2023bosonic}. In the future, implementing doublon detection~\cite{gross2021quantum, preiss2015quantum, hartke2020doublon, koepsell2020robust} could allow the study of correlations at $r = 0$, where we expect to see the strongest correlation signal, and possibly also allow the probing of interaction effects that are negligible on the length scales in this study. Such interaction effects become dominant in a Tonks gas \cite{hao2022n}, which could also be studied \textit{in situ} with our technique. 

\begin{acknowledgments}
While this work was in progress, we became aware of related work on \textit{in situ} studies of correlations in ultracold Fermi and Bose gases~\cite{yefsah_arxiv, zwierlein_private_comm}. 

We acknowledge Woo Chang Chung for early contributions to the design of the quantum gas microscope, Hanzhen Lin for experimental assistance, and Yoo Kyung Lee and Yu-Kun Lu for critical reading of the manuscript. We acknowledge support from the NSF through grant No. PHY-2208004, the Center for Ultracold Atoms (an NSF Physics Frontiers Center) through grant No. PHY-2317134, and the Army Research Office (award number: W911NF-24-1-0218, grant No.~W911NF-22-1-0024, DURIP).
\end{acknowledgments}

\bibliography{bib.bib}

\clearpage
\setcounter{equation}{0}
\setcounter{figure}{0}
\setcounter{table}{0}
\setcounter{page}{1}
\makeatletter
\renewcommand{\theequation}{S\arabic{equation}}
\renewcommand{\thefigure}{S\arabic{figure}}
\renewcommand{\bibnumfmt}[1]{[S#1]}
\renewcommand{\citenumfont}[1]{S#1}

\onecolumngrid

\section*{Supplemental materials}

\subsection*{Experimental apparatus and sequence}

\begin{center}
\begin{figure}[h]
    \includegraphics[width=0.8\textwidth]{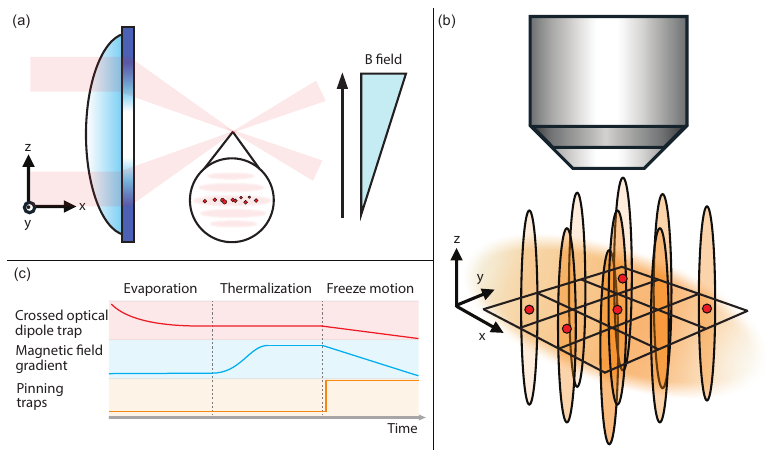}
    \caption{Experimental setup. (a) A 12-$\mu$m spacing vertical lattice is formed by intersecting two 1064-nm beams, which are focused onto the atoms at a 5$^{\circ}$ intersection angle by an aspheric lens. A magnetic field gradient is applied in the vertical direction to compensate for gravity. 
    (b) To image the atom positions, the atoms are pinned down by two horizontal lattices intersected at approximately 90$^{\circ}$. This forms a square grid of tubes elongated in the z-direction with 532-nm spacing. A light sheet, depicted as a diffuse oval along the x-axis, provides tight confinement in the z-direction. We employ polarization gradient cooling and collect the fluorescence photons with a microscope objective. 
    (c) The experimental sequence is graphically depicted. The atoms are cooled in the crossed dipole trap by lowering the trap depths over 10 seconds. We subsequently ramp up a magnetic field gradient over 500 ms to compensate for gravity. To freeze the spatial distribution, we quench the pinning lattices in sub-microsecond time and ramp down the crossed dipole trap and magnetic field gradient. Fluorescence detection is performed in the deep pinning lattices.}
    \label{fig:experiment}
\end{figure}
\end{center}

Schematics and key steps of the experimental sequence to prepare and characterize a quasi two-dimensional thermal gas are shown in Fig.~\ref{fig:experiment}. The experimental machine is described in detail in Ref.~\cite{streed2006large}.

We start with approximately $10^4$ $^{87}$Rb atoms prepared in the spin-polarized $|F=1, m_F=-1\rangle$ state loaded into a single node of a 12-$\mu$m spacing vertical lattice, as depicted in Fig.~\ref{fig:experiment}(a). The vertical lattice node has a cross-section with Gaussian widths of 5.4 $\mu$m and 96 $\mu$m in the vertical and horizontal directions, respectively. The atoms are loaded into the lattice from a co-propagating light sheet with Gaussian widths of 8 $\mu$m and 36 $\mu$m. An additional laser beam propagating in the y-direction, not depicted in Fig.~\ref{fig:experiment}(a), is used to provide confinement along the vertical lattice beam's propagation direction to form a crossed optical dipole trap. The atoms are then cooled through forced evaporation by lowering the trap depth of the crossed dipole trap over 10 seconds to allow for thermalization through collisions. A magnetic field gradient in the vertical direction is ramped up over 500 ms to compensate for gravity and ensure a favorable vertical trap frequency. The final trap frequencies for a sample with $\approx$ 100 atoms and $T \approx$ 6.5 nK are $(\omega_{x}, \omega_{y},  \omega_{z}) = 2 \pi \times (12, 15, 380)$ Hz. We measure the trap frequencies by parametric heating. 

To observe correlations in the thermal gas, we carefully control the total atom number $N$ to ensure that the BEC phase transition temperature $T_\mathrm{c}$ is lower than $T$.

For detection, we freeze the atomic motion by quenching on two horizontal pinning lattices intersected at 90$^{\circ}$ to $3000 \; E_{R}$ in sub-microsecond time before ramping down the cross dipole trap and magnetic field gradient (Fig.~\ref{fig:experiment}(b)). The light sheet is subsequently switched on to a few thousand $E_{R}$, where $E_{R} = h^2/(2m\lambda^2)$ is the recoil energy of a lattice photon with wavelength $\lambda =$ 1064~nm for an atom of mass $m$. The atomic spatial distribution is projected onto a square grid with 532-nm spacing. The experimental sequence is summarized in Fig.~\ref{fig:experiment}(c). We then measure the lattice occupation by applying polarization gradient cooling and collecting the fluorescence photons with a microscope objective. We collect $\approx$ 1000 photons per atom over 1000 ms. Our microscope objective has an NA = 0.8, allowing for a measured point spread function with a FWHM of 640~nm. It sits outside of the vacuum system on a 5 mm-thick fused silica viewport, which, in turn, is 6.5 mm away from the atoms. The objective has an effective focal length of 30 mm. An eyepiece with a focal length of 2 m focuses the image onto a camera with pixel size 13 $\mu$m. We calibrate the magnification to be a factor of 62.

\subsection*{Image reconstruction}

\begin{center}
\begin{figure}[h]
    \includegraphics[width=\textwidth]{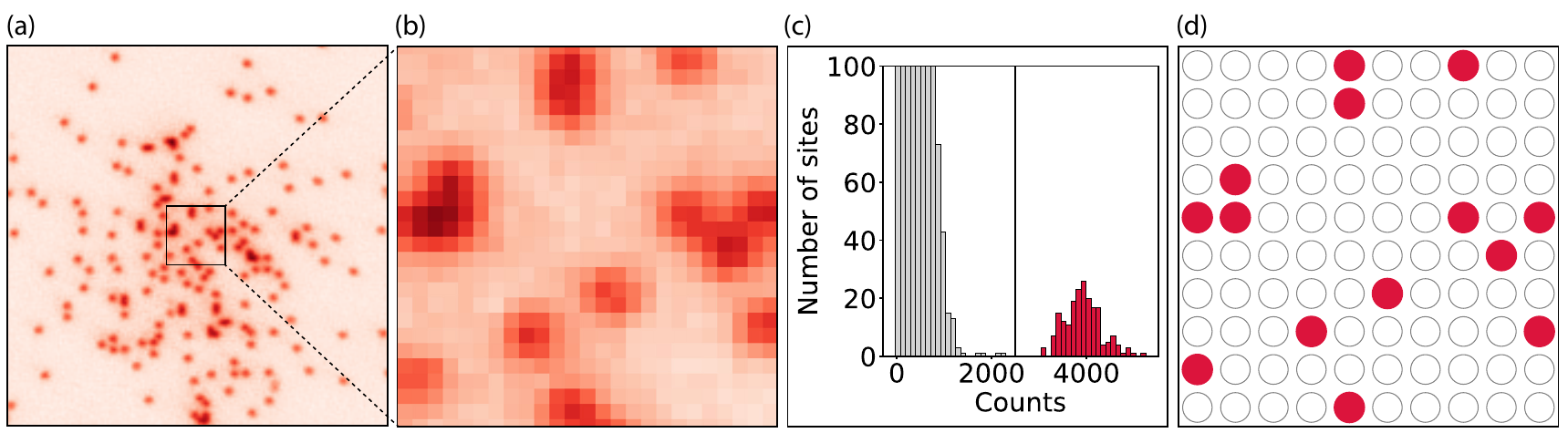}
    \caption{Reconstruction procedure. (a) A typical raw, dilute fluorescence image with good signal-to-noise is shown. Atomic fluorescence is collected by our microscope objective, which provides high spatial resolution for imaging single atoms. (b) We show a magnified view of a small region of the raw image, depicted for illustration purposes. A single atom with a point spread function with a FWHM of $\sim 640$~nm is distributed over multiple camera pixels. (c) We bin pixel counts on a given lattice site and then generate a histogram showing the number of lattice sites with a given count value. The majority of lattice sites are unoccupied since the sample is dilute, as shown in the left peak of the histogram. The occupied lattice sites are shown in the right peak of the histogram in red. Based on this histogram, we determine a threshold (black vertical line), above which lattice sites are considered occupied. (d) We depict the digitized version of lattice site occupation, termed the ``reconstructed" image. Lattice sites filled in red are occupied, while unshaded circles are unoccupied lattice sites. This occupation matrix is used for the second-order correlation analysis.}
    \label{fig:reconstruction}
\end{figure}
\end{center}

The process of reconstructing lattice site occupations is outlined in Fig.~\ref{fig:reconstruction}. Raw fluorescence images are deconvoluted with a regularized inverse filter based on the atomic point spread function, and the lattice spacing and angle are calibrated. Then, camera pixel counts on each lattice site are binned to produce a histogram of counts, as shown in Fig.~\ref{fig:reconstruction}c. A threshold is chosen, above which lattice sites are deemed to be occupied. The reconstructed lattice site occupations are then used for the analysis of the second-order correlations presented in this work.

\subsection*{Temperature measurement}

\begin{center}
\begin{figure}[tb]

  \includegraphics[width=\textwidth]{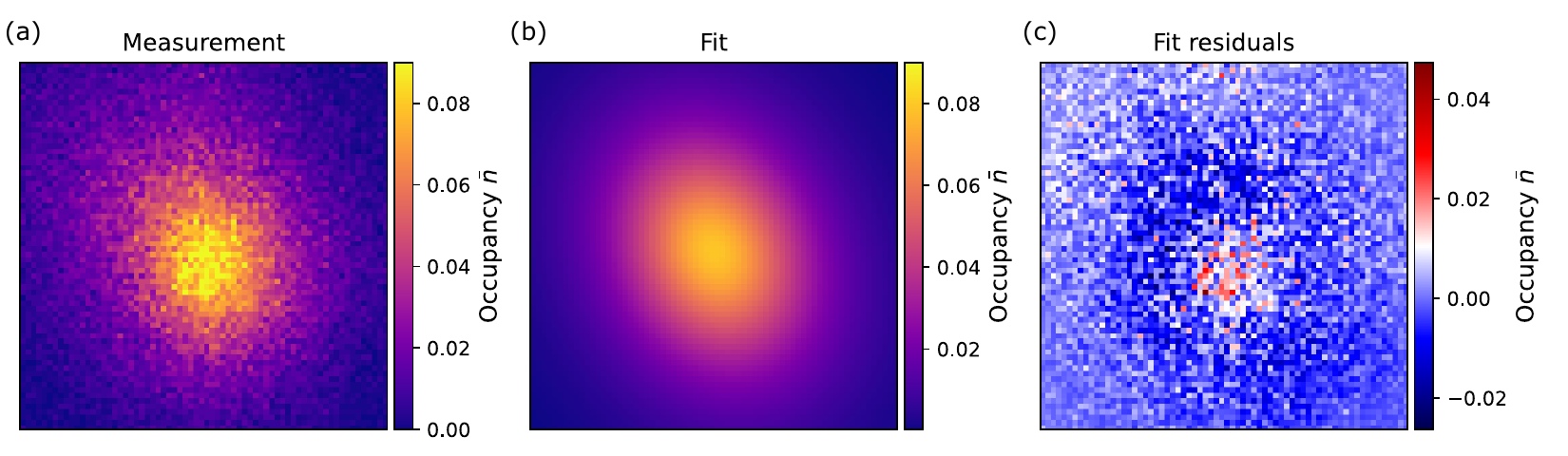}
\caption{Temperature determination. (a) The mean occupancy $\bar{n} (x, y)$ on each lattice site averaged over all ($\approx 650$) experimental snapshots. (b) The fit to the data in panel (a) using Eq.~(\ref{eq:density_dist}). We extract $T = 6.5$ nK. (c) Residuals of fit. }
    \label{fig:s1}

\end{figure}
\end{center}

For our system size ($N \approx 100$), finite-N correction terms for the thermodynamic equations of Bose gases (generally calculated for $N \rightarrow \infty$) are relevant. Following the treatment in Ref.~\cite{ketterle1996bose}, we properly estimate the populations in all oscillator states.  We begin with the Bose-Einstein distribution formula:
~
\begin{equation}
\label{eq:BE_dist}
    n_{k}(E_{i}, T) = \frac{1}{e^{\beta({E_{k}} - \mu )} - 1}. 
\end{equation}

For the three-dimensional harmonic trap, the energy for a given mode $k$ is $E_{k} = \hbar (\omega_{x} n_{x} + \omega_{y} n_{y} + \omega_{z} n_{z} + 3/2)$.  We count all states so no degeneracy factors are required. The chemical potential $\mu$ is treated as a normalization factor to ensure $N = \sum_{k} n_{k}(E_{k})$. 

First, we model the temperature of our ensemble including finite size effects. We average many experimental snapshots to determine the mean occupancy per lattice site $\bar{n} (x, y)$ as plotted in Fig.~\subref{fig:s1}{a}. Next, we fit the density distribution as a sum of Hermite polynomials denoted $\psi_{n_{x}, n_{y}} (x, y)$ (the solutions to the quantum harmonic oscillator):
~
\begin{equation}
\label{eq:density_dist}
    n_{\text{fit}} (x, y) = \sum_{n_{x}}  \sum_{n_{y}} n_{k} (n_{x}, n_{y}, T) |\psi_{n_{x}, n_{y}} (x, y)|^{2} . 
\end{equation}
~
$\psi_{n_{x}, n_{y}} (x, y)$ depends on the trap frequencies $\omega_{x} , \omega_{y}$ which are independently calibrated via parametric heating. We fit $T = 6.5$ nK and plot the fitted density distribution in Fig.~\subref{fig:s1}{b}. Parameterized in terms of the two-dimensional condensate temperature in the thermodynamic limit ($N \rightarrow \infty$),
\begin{equation}
    k_{B} T_\mathrm{c}^{2D} = \sqrt{N} \hbar \bar{\omega}_{R}\sqrt{6/\pi^{2}}, 
\end{equation}

\noindent we find $T / T_\mathrm{c}^{2D} = 1.26$. The fit residuals are plotted in Fig.~\subref{fig:s1}{c}, showing good agreement between the experiment and data. Fitting instead to a two-dimensional Bose gas in the thermodynamic limit, we obtain $T = 5.6$~nK in reasonable agreement.

The effects of interactions are not included in this fit since they are weak.  The first order correction would be adding a mean-field term to the Hamiltonian of $U_{\text{MF}} =  \tilde{g} \mathcal{D}(r) / \pi$~\cite{hadzibabic2011two}. $\mathcal{D}(r)$ is the local phase space density. The dimensionless interaction strength is $\tilde{g} = \sqrt{8 \pi} a / l_{z}$ with the tight-confinement harmonic oscillator length $l_{z} = \sqrt{\hbar / m \omega_{z}}$. For our trap parameters and s-wave scattering length $a \approx 100~a_{0}$, $\tilde{g} = 0.04$, and thus interactions are a small correction to the extracted temperature $T$. 

\subsection*{Finite system size corrections}

The finite $N \approx 100$ size of the system also requires a correction of the $g^{(2)}$ function since low-lying states no longer have small occupation numbers.  To estimate this effect, we model our system as homogeneous (a box potential) and study the $g^{(2)}$ correction. The $g^{(2)}$ function can be expressed at $ r = 0$ as~\cite{walraven_notes}:

\begin{equation}
    \label{eq:g2_walraven}
    G^{(2)}(r = 0) = \frac{1}{V^{2}} \sum_{k \neq l} 2 n_{k} n_{l} + \frac{1}{V^{2}} \sum_{k} n_{k} (n_{k} - 1) . 
\end{equation}

Given $N^{2} = (\sum_{k} n_{k})(\sum_{l} n_{l}) = \sum_{k}n_{k}^{2} + \sum_{k \neq l} n_{k} n_{l}$, rearranging terms we arrive at:

\begin{equation}
\label{eq:g2_WK}
    g^{(2)}(r = 0) = 2 - \frac{1}{N} - \sum_{k} \langle n_{k}^{2} \rangle / N^{2}. 
\end{equation}

Eq.~(\ref{eq:g2_WK}) is exact for the canonical ensemble.  We  use  $\langle n^{2}_{k} \rangle = \langle n_{k} \rangle^{2} + \text{Var}(n_{k})$.   There is no explicit expression for $\text{Var}(n_{k})$ in the canonical ensemble.  However, for sufficiently small population numbers, one can regard each state in a bath of the atoms in other states, and use the grand-canonical expression $\text{Var}(n_{k}) = \langle n_{k} \rangle +\langle n_{k} \rangle^{2}$.

\begin{equation}
\label{eq:g2_GCE}
    g^{(2)}(r = 0) = 2 - \frac{2}{N} - 2\sum_{k} \langle n_{k} \rangle^{2} / N^{2} . 
\end{equation}

We numerically solve for $\langle n_{k} \rangle$ for a range of different $N$ values at $T = T_\mathrm{c}$. For a three-dimensional homogeneous potential, the critical temperature is:

\begin{equation}
\label{eq:3D_homogeneous}
    k_{B} T_\mathrm{c}^{3D} = \frac{4 \epsilon_{1}}{\pi} \Big( \frac{N}{\zeta(3/2)} \Big)^{2/3}. 
\end{equation}

\begin{figure}[!htbp]
    \centering
    \includegraphics[width=0.475\textwidth]{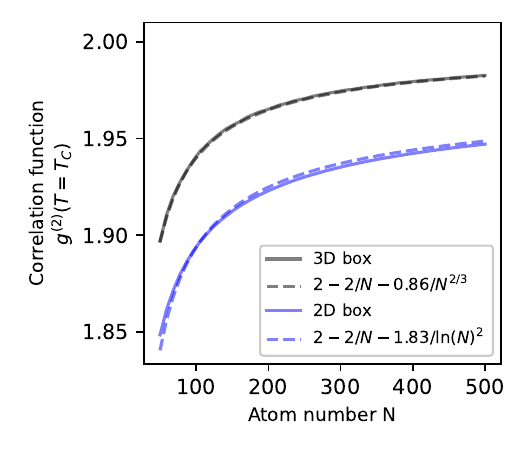}
    \caption{$g^{(2)}$ corrections. For a 2D and 3D homogeneous potential at $T = T_\mathrm{c}$, we vary the total atom number $N$ and numerically determine the state populations $ \langle n_{k} \rangle$ from a Bose-Einstein distribution. From Eq.~(\ref{eq:g2_GCE}), we calculate $g^{(2)}$.}
    \label{fig:numerical_g2_corrections}
\end{figure}

This is calculated in the thermodynamic limit using an integral with the density of states $N_\mathrm{c} = \int_{0}^{\infty} d \epsilon \; g(\epsilon) \langle n_{k} \rangle $, where $g(\epsilon) \propto \sqrt{\epsilon}$ for a 3D box.  Here $\zeta$ is the Riemann Zeta function and $\epsilon_{1} = \pi^{2} \hbar^{2}/2mL^{2}$ is the first excited state energy. At criticality ($T = T_\mathrm{c}$) in two and three dimensions, we systematically vary the total atom number $N$ and numerically calculate $g^{(2)}$ from Eq.~(\ref{eq:g2_GCE}). The numerically calculated $g^{(2)}$ function at $r = 0$ is plotted in Fig.~\ref{fig:numerical_g2_corrections}. 

To gain further intuition and to confirm our numerics, we derive an approximate analytical expression for the $g^{(2)}$ function. This requires obtaining an analytical expression for $2\sum_{k} \langle n_{k} \rangle^{2} / N^{2}$. These sums can first be written as integrals evaluated in a 3D homogeneous potential:

\begin{equation}
  \sum_{k} \langle n_{k} \rangle^{2} / N =  \int_{0}^{\infty} d \epsilon \; g(\epsilon) \langle n_{k} \rangle^{2} / \int_{0}^{\infty} d \epsilon \; g(\epsilon) \langle n_{k} \rangle .  
\end{equation}

\noindent The integral $\int_{0}^{\infty} d \epsilon \; g(\epsilon) \langle n_{k} \rangle$ is analytically solvable (as in Eq.~(\ref{eq:3D_homogeneous})) and equals $\zeta(3/2) \Gamma (3/2) = 2.61 \sqrt{\pi}/2$. 
The integral $\int_{0}^{\infty} d \epsilon \; g(\epsilon) \langle n_{k} \rangle^{2}$ diverges as $\epsilon \rightarrow 0$. To address this we set the lower bound of the integrand at the energy of the first excited state $\beta \epsilon_{1}$. 

\begin{equation}
   \frac{1}{\beta^{3/2}} \int_{\beta \epsilon_{1}}^{\infty} d \beta \epsilon \; \frac{\sqrt{\beta \epsilon}}{[e^{\beta \epsilon} - 1]^{2}} . 
\end{equation}

Applying a Taylor expansion $e^{x} \approx 1 + x$ to isolate the singular behavior, we arrive at $\epsilon^{-1/2} |^{\infty}_{\beta \epsilon_{1}}$. Re-writing $\beta \epsilon_{1} = \pi/4 [\zeta(3/2) / N]^{2/3}=1.49 N^{-2/3}$ and plugging in values, we obtain the key result:

\begin{equation}
    \label{eq:N_scaling}
    \sum_{k} \langle n_{k} \rangle^{2} / N \approx 0.71 N^{1/3} . 
\end{equation}

\noindent Plugging Eq.~(\ref{eq:N_scaling}) into Eq.~(\ref{eq:g2_GCE}), we obtain an analytical expression for the $g^{(2)}$ correlation function for a 3D homogeneous system. 

\begin{equation}
\label{eq:g2_GCE_analytical}
    g^{(2)}(r = 0) \approx 2 - \frac{2}{N} - 1.4 N^{-2/3}.  
\end{equation} 

The prefactor for the $N^{-2/3}$ term is only an approximation, since we replaced the discrete summation by an integral with a cutoff at $\epsilon_1$. Any other choice of cutoff will yield the same $N^{-2/3}$ term, albeit with a different prefactor.

Another derivation focuses on the lowest energy states. For the lowest energy state, the occupation number $n_0= -1/\beta \mu$, where $\mu$ is the chemical potential.  For the first excited state it is $n_1= 1/\beta (\epsilon_1 - \mu)$. To estimate the scaling of $\mu$ at finite particle number, we express the total atom number splitting off the ground state 

\begin{equation}
\label{eq:3D_box_mu}
N = \frac{L^{3}}{\lambda_{\text{dB}}^{3}} g_{3/2}(z) + \frac{z}{1-z} , 
\end{equation}

\noindent in terms of the fugacity $z = e^{\beta \mu} = e^{- \alpha}$. Expanding $z$ around $z = 1$ for small $\alpha$, $g_{3/2}(\alpha) \approx \zeta (3/2) - 2 \sqrt{\pi \alpha}$. Plugging into Eq.~\ref{eq:3D_box_mu}, $N = \frac{L^{3}}{\lambda_{\text{dB}}^{3}} [\zeta(3/2) - 2 \sqrt{\pi \alpha}] + \frac{1}{\alpha}$. Using $N_\mathrm{c} = \frac{L^{3}}{\lambda_{\text{dB}}} \zeta (3/2)$, we obtain $\alpha = [\frac{\zeta(3/2)}{2 \sqrt{\pi}}]^{2/3} N_\mathrm{c}^{-2/3} = 0.82 N_\mathrm{c}^{-2/3} = 0.55 \beta \epsilon_{1}$. Since $1/\beta \propto N^{2/3}$, one finds again an $N^{-2/3}$ correction term for $g^{(2)}$. Our numerical calculations confirm the $N^{-2/3}$ scaling with a fitted prefactor of $0.86$.

For a two-dimensional homogeneous system, the density of states is constant. Thus $N_\mathrm{c} = \int_{0}^{\infty} d \epsilon \langle n_{k} \rangle $ diverges. To address this, the first excited state $\epsilon_{1}$ may be used as described in Ref.~\cite{ketterle1996bose}. Doing this, one arrives at an expression for the effective critical temperature  $\tilde{T}_\mathrm{c}^{2D}$: 

\begin{equation}
N = \frac{ k_{B} \tilde{T}_\mathrm{c}^{2D} \pi }{4 \epsilon_{1}} \ln{ \Big( \frac{ k_{B} \tilde{T}_\mathrm{c}^{2D} \pi }{4 \epsilon_{1}} \Big) } , 
\end{equation}
or in leading order
\begin{equation}
\frac{ k_{B} \tilde{T}_\mathrm{c}^{2D} \pi }{4 \epsilon_{1}} = N/\ln N . 
\end{equation}

Using the same approximations as in 3D, one finds the population of the low-lying states to be proportional to $\frac{1}{\beta \epsilon_1} $.  But now $1/\beta \propto N / \ln N$, and therefore the correction term for $g^{(2)}$ scales with $1/(\ln N)^2$.  Numerical calculations (see Fig.~\ref{fig:numerical_g2_corrections}) are in agreement with this predicted scaling. For a 2D homogeneous box at our operating temperature $T / T_\mathrm{c}^{2D} = 1.26$ and atom number, we obtain a $g^{(2)}$ value of $1.93$. 

Note that the grand canonical ensemble gives the unphysical result $g^{2}(0) = 2$, even below the transition temperature and also for a pure condensate. For the grand canonical ensemble, Eq.~(\ref{eq:g2_WK}) has an extra term $\text{Var}(N)/N^2$ which exactly cancels the finite-$N$ corrections. It is well-known that $\text{Var}(N)$ becomes unphysical for ideal Bose gases in the grand canonical ensemble~\cite{wilkens1997particle, kocharovsky2006fluctuations}.

\subsection*{Finite spatial resolution}

As expressed in Eq.~(\ref{g2_eq}) in the main text, the $g^{(2)}$ correlation function is diminished by the finite resolution of the quantum gas microscope. As pointed out in Ref.~\cite{pyzh2019quantum}, the microscope has a ``quantum point spread function" limited by the fidelity of the pinning procedure. We qualitatively model this effect as a Gaussian blurring process. Blurring effects on the $g^{(2)}$ function from finite detector resolution have already been modeled for previous experiments in Ref.~\cite{gomes2006theory}. In the absence of atoms in the BEC state, the $g^{(2)}$ function can be rewritten as the un-normalized $G^{(1)}$ function:
~
\begin{equation}
    g^{(2)}(0, 0) = 1 +  \frac{|G^{(1)}(0, 0)|^2}{\rho(0)^2}. 
\end{equation}
\noindent
Next, we express the $G_{1}$ correlation function as a Gaussian

\begin{equation}
  G^{(1)}  =   \exp{(-\pi\frac{r^2}{\lambda_{\text{dB}}^2})} = \exp{(-\frac{r^2}{4l^2})}, 
  \\ 
\end{equation} 

\noindent parameterized by a correlation length $l = \lambda_{\text{dB}}/(2\sqrt{ \pi})$, which is the Gaussian standard deviation for the $G^{(2)}$ function. The correlation length is substantially shorter than $\lambda_{\text{dB}}$. 

Next closely following the derivation from Ref.~\cite{gomes2006theory}, we determine how finite spatial resolution affects the experimentally observed correlation function $G_{\text{obs}}^{(1)}$. The effective imaging resolution is modeled by convoluting  $G^{(1)}$ with a Gaussian of standard deviation $\sigma$: 
~
\begin{align}
    |G_{\text{obs}}^{(1)}(x, x^{\prime})|^{2} &= \int dx_{0} dx_{0}^{\prime} |G^{(1)}(x_{0}, x_{0}^{\prime})|^{2}
    \times \frac{e^{-(1/2) [(x-x_{0})/2\sigma]^{2}}}{\sqrt{2 \pi }\sigma} \frac{e^{-(1/2) [(x^{\prime}-x^{\prime}_{0})/2\sigma]^{2}}}{\sqrt{2 \pi }\sigma} \\
    &=  \frac{|A|^{2}}{\sqrt{1 + 2 \sigma^{2}/l^{2}}} e^{-(x - x^{\prime})^{2}/[2(l^{2} + 2\sigma^{2})]}. 
\end{align}

\noindent
Evaluated at $(x = 0, x^{\prime} = 0)$ we have:
~
\begin{equation}  g_{\text{obs}}^{(2)}(0, 0)-1 = \frac{|G_{\text{obs}}^{(1)}(0, 0)|^2}{\rho_{\text{obs}}(0)^2} = \prod_{\alpha = x,y}\frac{1}{\sqrt{1+2\sigma_\alpha^2/l_\alpha^2}} = \frac{l^2}{2\sigma^2+l^2}. \end{equation}

\noindent Using our experimental parameters $\lambda_{\text{dB}} = 2.32 \; \mu \text{m}$, $l_{\alpha} = 1.23 \times 532 \; \text{nm}$, and fitted $\sigma = 1.25 \times 532 \;  \text{nm}$ we find $\frac{l^2}{l^2 + 2\sigma^2} = 0.33$ when performing a 2D convolution.

\subsection*{Atom number fluctuation}
For the cold ensemble, we average over approximately 650 images to determine the probability of detecting one particle per site, denoted by $\langle p\rangle$, where $\langle \cdot \rangle$ represents the ensemble average over experimental realizations. When $p$ varies from shot to shot through total atom number fluctuations, the measured number of pairs will be enhanced due to the Cauchy-Schwarz inequality: $\langle p^2\rangle \geq \langle p \rangle^2 $. Therefore, in the first analysis, when we normalize the number of observed pairs by  $\langle p\rangle^2$, we obtain an enhanced $g^{(2)}$ function. However, this effect does not depend on the pair separation distance $r$, it only contributes an overall scaling factor to the absolute value of $g^{(2)}(r)$ across all pairs. 
This was verified by numerical simulations of classical particles with added atom number fluctuations.  We confirmed that such fluctuations indeed only contribute an overall scaling factor that is independent of pair separation.

Therefore, in the data analysis, we normalize $g^{(2)}(r)$ by the pairs separated at a distance larger than 3 $\mu$m where density fluctuations are uncorrelated. We also tried to post-select data based on total atom number and observed very similar results.

\subsection*{Spatial inhomogeneity}
In the experiment, we determine the single particle filling probability for a region to be $\bar{p}$, where the average is taken over a spatial region. Within this region, consider a pair located at ($\textbf{r}_1$, $\textbf{r}_2$) with corresponding local single-particle filling probabilities of $p_1$ and $p_2$ which are slightly different from $\bar{p}$ due to spatial inhomogeneity. The measured number of those pairs is proportional to $p_1p_2$, whereas the number of uncorrelated pairs with short separation at positions $r_1$ and $r_2$ is proportional to $\bar{p^2} = \frac{p_1^2+p_2^2}{2}$.
The inequality $(p_1-p_2)^2\geq0$
implies $\frac{p_1^2+p_2^2}{2} \geq p_1 p_2$.  This means that even for classical particles, the number of pairs will be smaller for larger pair separation.

This affects the data analysis because we use the pairs separated at a distance larger than 3 $\mu$m to normalize the absolute value of $g^{(2)}(r)$.

To address this issue, as described in the main text, we segment the total field of $70 \times 70$ sites into $49$ smaller $10 \times 10$ boxes, each with a different local density which is now almost homogeneous within the box. We then plot the measured number of pairs against the expected number of pairs in each box and fit a line $y = ax$ to the data. The slope of this line is the $g^{(2)}$ value for the specific pair separation $r$. Despite this approach, there remains approximately 10\% residual spatial inhomogeneity for pairs separated by 5 $\mu$m within each $10 \times 10$ box. In the presence of a density gradient, the correction between $\bar{p^2}$ and $p_1p_2$ simply varies quadratically as the pair separation $r$ increases:
~
\begin{equation}
     \frac{p_1p_2}{(p_1^2+p_2^2)/2} = \frac{1 - \Delta^2}{1 + \Delta^2}
    \approx  1 - 2 \Delta^2
    = 1 - c r^2
\end{equation}
\noindent
where $p_1=\bar{p}(1+\Delta)$,   $p_2=\bar{p}(1-\Delta)$
and  $\Delta \propto r$. 

\begin{figure}[!htbp]
    \centering
    \includegraphics[width=0.6\textwidth]{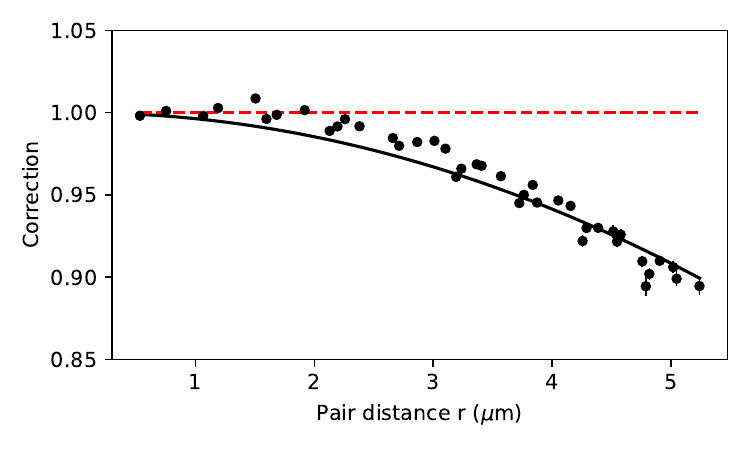}
    \caption{Correction for spatial inhomogeneity. For classical particles drawn from a uniform spatial density distribution, the $g^{(2)}$ function should measure 1 for all pair distances (red dashed line). Due to residual spatial inhomogeneity within each $10\times10$ box, the pair-counting method used to determine $g^{(2)}(r)$ will provide a systematically smaller number of pairs at large distances. This requires a correction of approximately 10\% for pairs separated by 5~$\mu$m. The figure shows the number of pairs of numerically simulated classical particles drawn from the experimentally measured density distribution (black dots). The error bars represent the standard error of the mean across 100 independent numerical simulations. The number of pairs is normalized to one at zero distance. The behavior is well captured by a quadratic function, $y = 1 - c r^2$, where $c$ is a fitting parameter (black solid line).}
    \label{fig:spatial_inhomo}
\end{figure}

To validate this correction, we conducted numerical simulations of classical particles drawn from the experimentally measured density distribution. Feeding the simulated data, which we expect to yield $g^{(2)}(r) = 1$, into our data analysis pipeline confirmed this quadratic correction, in agreement with the prediction above from the simple linear gradient model. The results are summarized in Fig.~\ref{fig:spatial_inhomo}. We observed similar behavior in the initial data analysis of the hot sample where $g^{(2)}(r)\approx1$, confirming the need for this classical systematic correction. 

\subsection*{Correction factor for finite vertical confinement}
Atoms in different $n_z$ states of the vertical harmonic oscillator potential are ``distinguishable" and therefore uncorrelated, reducing the $g^{(2)}$ contrast. We know the state populations based on the thermometry. The fraction of atoms in a $n_{z}$ level is denoted $P_{n_{z}}$. We calculate $P_{0} = 0.95$ using Eq.~(\ref{eq:BE_dist}) with the fitted experimental temperature. We note that our system is thermodynamically well approximated by a Maxwell-Boltzmann distribution ($P_{0}^{MB} = 0.94$).

We evaluate the first three z levels and calculate the correction:
~
\begin{equation}
\label{eq:eta_nz}
    \eta = P_{0}^{2} + P_{1}^{2} + P_{2}^{2} = 0.91
\end{equation}

\end{document}